\documentclass[a4paper,nofootinbib,showpacs,11pt]{revtex4}

\usepackage{amssymb,amsmath}
\usepackage{latexsym}
\usepackage{graphicx}
\usepackage{amsfonts}
\usepackage{amssymb}
\usepackage{epsfig}
\usepackage{color}
\usepackage{psfrag}

\def\lt#1{\left#1}
\def\rt#1{\right#1}

\def\frc#1#2{\frac{#1}{#2}}

\newcommand{\p}{\partial}

\newcommand{\bra}{\langle}
\newcommand{\ket}{\rangle}
\newcommand{\Z}{{\mathbb{Z}}}

\begin{document}

\title{Energy flow in non-equilibrium conformal field theory}

\author{Denis Bernard${}^{\clubsuit}$\footnote{Member of C.N.R.S.; \texttt{denis.bernard@ens.fr}}
and Benjamin Doyon${}^{\spadesuit}$\footnote{\texttt{benjamin.doyon@kcl.ac.uk}} }

\date{\today} 

\affiliation{
${}^{\clubsuit}$ Laboratoire de Physique Th\'eorique, CNRS $\&$ Ecole Normale Sup\'erieure de Paris, France.\\
${}^{\spadesuit}$ Department of Mathematics, King's College London, London, United Kingdom. }

\begin{abstract}
We study the energy current and its fluctuations in quantum gapless 1d systems far from equilibrium modeled by conformal field theory, where two separated halves are prepared at distinct temperatures and glued together at a point contact. We prove that these systems converge towards steady states, and give a general description of such non-equilibrium steady states in terms of quantum field theory data. We compute the large deviation function, also called the full counting statistics, of energy transfer through the contact. These are universal and satisfy fluctuation relations. We provide a simple representation of these quantum fluctuations in terms of classical Poisson processes whose intensities are proportional to Boltzmann weights.
\end{abstract}


\maketitle

{\it Introduction.}
A lot of experimental and theoretical progress has been achieved in non-equilibrium physics over the past years, see for instance \cite{zwanzig,blanter}. A popular set of tools and ideas in the classical realm are the classical fluctuation relations \cite{fluctu} and large deviation techniques \cite{Kurchan,Lebow}, which led to the understanding of universal properties of far-from-equilibrium systems. Elements of fluctuation theory has been extended to quantum systems, see e.g.~\cite{deRoeck,Espo}, hoping this will likewise reveal principles governing non-equilibrium quantum physics. In particular, a lot of attention has been given to mesoscopic electronic systems in which a steady state far from equilibrium exists, where a current (of charge, energy, etc.) is flowing between various parts. In these systems, most interesting are current fluctuations and their full counting statistics (FCS). The full counting statistics encompasses the statistics of current fluctuations, and is usually encoded into a formula generating the leading behavior of all the cumulants of the observable measuring the quantity transferred after a long time. By opposition to the classical situation, this observable fluctuates not only thermally, but also quantum mechanically. For non-interacting charged-fermion systems, the full counting statistics of the electric current is given by the celebrated Levitov-Lesovik formula \cite{Lev1}. Further understanding into electric charge transfer has been gained for some low-dimensional interacting systems using bosonization \cite{Mirlin} or Bethe ansatz techniques \cite{Saleur}. Here we extend this progress by analyzing the energy current: we determine the large-time cumulant generating function, simply related to the large deviation function, for the energy current in any critical quantum one-dimensional system (with dynamical exponent $z=1$) in a steady state far from equilibrium, and we show its universal character. As far as we know, there is currently only one other known exact generating function for energy current fluctuations: the case of a chain of harmonic oscillators \cite{SaitoDhar}, which, in its universal scaling limit, reproduces our result.

Although we will specialize to one dimension, let us start more generally: consider a quantum system with degrees of freedom lying on a d-dimensional lattice and interacting locally (few-neighbors interactions). Suppose that the system is initially prepared whereby two halves of it, say its left $(x<0)$ and right $(x>0)$ parts, are thermalized independently at different temperatures $T_l$ and $T_r$, and then glued together. Let it evolve for a very large time $t_o$. If the system is very large, in such a way that the distance to its extremities from the interface $x=0$ is much larger than the distance travelled in a time $t_o$ by the disturbance due to the gluing, then a stationary regime should take place. The system is then in a non-equilibrium steady state with energy transfer across the interface (in absence of translational or additional degrees of freedom, this energy transfer may be identified with the thermal energy transfer). 
We wish to describe {\it all} cumulants (including the average) of the large-time energy transfer in this steady state.

In order to study energy transfer, it is sufficient to consider effective degrees of freedom in terms of which the system is described more efficiently. There is a situation where these effective, collective degrees of freedom have a simple description: when the system is at, or near to, a critical point with unit dynamical exponent. Hence, let the correlation length and the size of the system be very large in lattice spacings, and the temperatures of the order of the corresponding energy gap. The result is described by a relativistically-invariant (massive) quantum field theory (QFT), and the effective degrees of freedom are the asymptotic particles. We recall that any given  QFT model is universal: it describes the near-critical behavior of every microscopic system in the same universality class. Let $h$ and $\vec{p}$ be the energy and momentum densities. They satisfy $\p_t h + \vec\nabla\cdot\vec{p}= 0$. The quantity whose fluctuations we want to analyze is the variations of the energy in one of the two halves, say $\tilde Q = \int_{x<0} d^dx\,h(x)$. The negative of its time variation is the integral of the momentum density perpendicular to the interface,
$-\p_t \tilde Q = \int_{x=0} d^{d-1} x \, p^\perp(x)$. In infinite volume this has infinite average if $d>1$, but the quantity $J=\bra p^\perp\ket$ is finite: this is the mean energy current per unit of transverse area.

After letting the system evolve during the time interval $[-t_o,0]$, its density matrix is $e^{-i\frac{t_o}{\hbar}H}\,\rho_0\, e^{i\frac{t_o}{\hbar}H}$ with $\rho_0$ the initial density matrix of the two thermalized halves and $H$ the system's hamiltonian. The steady state $\rho_{\rm stat}$ is obtained by sending $t_o$ to $+\infty$. Since $\rho_0$ is stationary with respect to the hamiltonian $H_o$ of the two decoupled halves, $\rho_{\rm stat}=S\, \rho_0\, S^{-1}$ with $S:=\lim_{t_o\to\infty}e^{-i\frac{t_o}{\hbar}H}e^{i\frac{t_o}{\hbar}H_o}$. That is: the $S$-matrix intertwines the initial thermalized state and the non-equilibrium steady state \cite{Ruelle}. In massive QFT, the Hilbert space of asymptotic particles is generally a product space ${\cal H}_+\otimes {\cal H}_-$ where ${\cal H}_\pm$ are spanned by states with particles going, respectively, towards the right $(p^\perp>0)$ or the left ($p^\perp<0$). Asymptotic states with positive transverse momenta come from free particles that were on the left in the far past, and vice versa. Within this picture we expect that the steady state density matrix factorizes, $\rho_{\rm stat}=\rho_+\otimes\rho_-$, and is diagonal in the basis of asymptotic particles. Its eigenvalues, on a state with particles at momenta $\vec{p}_j$ and energies $E_j$, are equal to $e^{-\beta_l \sum_{p_j^\perp>0} E_j}\,e^{ -\beta_r \sum_{p_j^\perp<0}  E_j}$ with $\beta^{-1}_{l,r}=k_BT_{l,r}$.

The main idea that emerges from the analysis above is that the density matrix for a thermal-flow steady state should be described simply if we know the right-moving and left-moving collective degrees of freedom (this is particularly clear in integrable models). Hence, although one may question the existence of the $S$-matrix in massless, or even scale-free, theories, we expect that in general, the steady state factorize on asymptotic left/right movers, in such a way that they are thermalized at different temperatures; essentially, these left/right movers are prepared in the far past in the asymptotic regions of the system which serve as effective reservoirs.

This picture can be made rather precise in one-dimensional exactly critical systems\footnote{For instance, any $z=1$ gapless quantum spin chain, like the Heisenberg chain, or the Ising chain at critical magnetic field.}, where any mass gap is much smaller than the temperatures (which are still much smaller than microscopic energy scales). In the scaling limit, these are described by $1d$ conformal field theory (CFT). There,  $p^\perp$ is a sum of left/right movers: $p^\perp_+\propto T_{zz}$ and $p_-^\perp\propto-{T}_{\bar z \bar z}$ with $T_{zz},\, {T}_{\bar z \bar z}$ the (anti)-holomorphic components of the stress tensor. Hence, the mean heat current is the sum of left/right contributions, $J=J_++J_-$ with $J_\pm =\bra p^\perp_\pm\ket$. This implies $ J = j(\beta_l) - j(\beta_r)$, since the energy of a particle is invariant under change of sign of $p^\perp$. Since there is no scale, dimensional analysis then tells us that $J \propto (T_l^{2} - T_r^{2})$. We calculate the mean energy current and find the universal formula\footnote{In the case $c=1$ and $T_r=0$, this formula bears similarities with the Stefan-Boltzmann law for the energy radiated by a thermal black body, see \cite{Cardy10}. We thank J. Cardy for pointing out this analogy.}:
\begin{eqnarray} \label{heatJ}
J = \frac{c\pi}{12\hbar}\, k_B^2(T_l^2-T^2_r).
\end{eqnarray}
This formula only depends on one parameter of the universality class: $c$, the CFT central charge\footnote{For non-unitary theories, the central charge $c$ in eq.(\ref{heatJ}) has to be replaced by $c_{\rm eff}=c-24 h_{\rm min}$ with $h_{\rm min}$ the minimal conformal dimension.}. For small temperature difference, $T_{l,r}=T\pm\Delta T/2$ with $\Delta T\ll T$, the mean energy current is $J=K\, \Delta T$ with thermal conductance $K=\frac{c\pi}{6\hbar} k_B^2 T$, as derived for a free fermions theory in \cite{fazio}. Such a formula for $K$ was shown in \cite{cappelli} to arise from the gravitational anomaly for CFT. Formula (\ref{heatJ}) appeared in \cite{Cardy2} in the different context of an inhomogeneous quantum quenches.

More interestingly, we also compute the full probability distribution of energy transfer during a large time $t$ (the energy full counting statistics). This is conveniently coded in the Legendre transform of the large deviation function\footnote{In the rest of the paper, by abuse of language, we will refer to this simply as the large deviation function.}, $F(\lambda):=\lim_{t\to\infty} t^{-1}\log\bra e^{i\lambda \Delta_tQ} \ket$ with $\Delta_tQ$ the energy transferred across the interface during time $t$. Our result is:
\begin{eqnarray} \label{energyFCS}
F(\lambda)= \frac{c\pi}{12\hbar}\, \Big( \frac{i\lambda}{\beta_r(\beta_r-i\lambda)}-\frac{i\lambda}{\beta_l(\beta_l+i\lambda)}\Big).
\end{eqnarray}
It is also very universal, depending only on the CFT central charge and universal constants. It satisfies the fluctuation relation \cite{fluctu,Espo},
\begin{eqnarray}\label{fluctu}
F(i(\beta_l-\beta_r)-\lambda)=F(\lambda)
\end{eqnarray}
That the energy transport fluctuations satisfy the fluctuation relation has been checked in the Pauli-Fierz model \cite{deRoeck2} and in the quantum harmonic oscillator chain \cite{SaitoDhar}. Although we are going to present a full field theory proof in \cite{BD12}, we will simply show here that eq.~(\ref{energyFCS}) is a consequence of the fluctuation relation, the fact that $F(\lambda)$ decomposes into the sum of left/right contributions, scale invariance, and the asymptotic behavior $F(\lambda)=-i\lambda\, J + o(\lambda)$ where $J$ is given by eq.~(\ref{heatJ}); we will assume the validity of the fluctuation relation. As usual, the fluctuation relation relates the probabilities $P_t(\theta)$ and $P_t(-\theta)$ of opposite energy transfers $\Delta_tQ=\pm t\theta$ across the interface:
\[ e^{-t\beta_l\theta}\, P_t(\theta)d\theta = e^{-t\beta_r\theta}\, P_t(-\theta)d\theta.\]
The large deviation function (\ref{energyFCS}) possesses a very natural interpretation, given below, in terms of Poisson processes whose intensities are proportional to Boltzmann weights and whose jumps are in correspondence with energy quanta, alias particles, crossing the interface. This leads us to propose natural generalizations for the energy FCS in massive (integrable) theories or including charge transfer.

{\it CFT out-of-equilibrium.}
Let us make the setting more precise. We use the standard Keldysh real-time construction of the steady state. We start with two {\it identical} gapless 1d quantum systems, each of length $R/2$, defined on intervals $[-R/2,0]$ and $[0,R/2]$, and prepared at respective temperatures $T_{l,r}$. We connect them through the origin at large negative time $-t_o$ so that the system state at time $0$, in any finite observation domain around the interface, is stationary for the coupled dynamics. The domain where there is a uniform and steady flow is of size of order $v_ft_o$. This has to be much smaller than the system size, because the extreme left and right parts away from this domain serve as effective thermal reservoirs, each at its own temperature $T_{l,r}$. Hence, we must have $R\gg v_ft_o\gg$ any observation or microscopic scales\footnote{Here $v_f$ is the typical excitation velocity. In the following we set $v_f=1$, $\hbar=1$, $k_B=1$.}. The steady state is mathematically defined by the limits $R\to\infty$ and then $t_o\to\infty$ in that order.

Before being connected, the two gapless systems are described by isomorphic CFT with central charge $c$. Let us recall here the standard results of CFT. The energy and momentum densities decompose as $h=h_+ +h_-$ and $p=h_+ - h_-$, with $h_\pm$ the chiral components (right- and left-moving respectively), $(\partial_t\pm\partial_x)h_\pm=0$. The boundary conditions are reflecting at all boundaries: $h_+(0^\pm,t)=h_-(0^\pm,t)$ and $h_+(\pm R/2,t)=h_-(\pm R/2,t)$, so that the system splits in its two independent left and right parts. Hence in each of the left ($x\in[-R/2,0]$) or right ($x\in[0,R/2]$) parts, we can safely set $h_\pm=h_\pm^{l,r}$, and results of CFT tell us that $h^{l,r}_+(x)=\frc{2\pi}{R^2} T_{R}^{l,r}(x)$ and $h^{l,r}_-(x)=\frc{2\pi}{R^2} T_{R}^{l,r}(-x)$ with $T^{l,r}_R$ the stress tensors of the left/right sub-systems. The stress-energy tensors have Fourier decompositions
\[ T^{l,r}_R(x):= -\frc{c}{24} + \sum_{n\in\Z}L_{n}^{l,r} e^{-2\pi i nx/R},\]
whose modes are Virasoro generators with commutation relations $[L_{n}^{l,r},L_{m}^{l,r}]=(n-m)L_{n+m}^{l,r}+ \frac{c}{12}n(n^2-1)\delta_{n+m;0}$. The hamiltonians $H^l_o = \int_{-R/2}^0 dx\, h^l(x)$ and $H^r_o= \int_{0}^{R/2} dx\, h^r(x)$ act respectively on the Hilbert spaces ${\cal H}^l$ and ${\cal H}^r$ (which are isomorphic as the sub-systems are identical).
Note that each sub-system, defined on an interval of length $R/2$, has been described as a periodic system with a single chiral component of the stress-energy tensor, but on a twice larger interval. This is a direct consequence of the conformal (i.e.~energy reflecting) boundary conditions.

After being connected, the system is still conformal, so that the energy and momentum densities still decompose into chiral components. The boundary conditions are now reflecting at the two extreme ends, $h_+(\pm R/2,t)=h_-(\pm R/2,t)$. At the contact point $x=0$, there are in principle many possibilities, corresponding to insertion of impurities. Here, however, we wish to describe a homogeneous system lying on $[-R/2,R/2]$, hence the conditions at $x=0$ are purely transmitting on each chiral component, $h_\pm(0^+,t)=h_\pm(0^-,t)$. The connected system is then described by chiral hamiltonian densities $h_+(x) = \frc{\pi}{2R^2} T_{2R}(x)$ and $h_-(x)=\frc{\pi}{2R^2} T_{2R}(R-x)$ acting on a Hilbert space ${\cal H}$. The hamiltonian is $H = \int_{-R/2}^{R/2} dx\,h(x)$. Clearly, we have ${\cal H}^l\otimes {\cal H}^r\hookrightarrow {\cal H}$. This map is implemented by the local identifications $h_\pm(x) = h_\pm^{l}(x)$ for $x\in[-R/2,0]$ and $h_\pm(x) = h_\pm^r(x)$ for $x\in[0,R/2]$ (here we use locality of both energy and momentum densities in order to separate $h_+$ from $h_-$) valid at the initial contact time.

Clearly, the difference between the $H_o$ and $H$-dynamics is on the boundary conditions. This may be rephrased as an abrupt change of conformally invariant defect, localized at the origin: before the contact time, the defect is factorizing, splitting the system in two parts, while after contact the defect is a so-called topological defect \cite{p-zuber}, letting the energy flow through\footnote{Conservation of energy for the total system imposes $(h_+-h_-)(0^-)=(h_+-h_-)(0^+)$. The stronger condition we impose amounts to assuming the absence of non-topological defects at the contact point.}. Topological defects include the absence of a defect (homogeneous system), but also certain defects making the system non-homogeneous but preserving the conformal symmetry. Our results hold for topological defects in general.

The stationary measure $\bra\cdots\ket_{\rm stat}$ may be viewed as a functional on operators of finite extent. We shall look at its action on the hamiltonian densities $h_\pm$. By definition $\bra \prod_j h_+^{(j)}(x_j) \prod_k h_-^{(k)}(y_k)\ket_{\rm stat}$ is equal to 
\begin{eqnarray} \label{statvev}
 \lim_{R\gg t_o\to \infty} \bra\prod_j h_+^{(j)}(x_j,t_o)  \prod_k h_-^{(k)}(y_k,t_o)\ket_0 
\end{eqnarray}
where $\bra\cdots\ket_0$ is the measure defined by the initial thermalized density matrix $\rho_0\propto e^{-\beta_l H_o^l}\otimes e^{- \beta_r H_o^r}$ and the time evolution is that of the coupled system, $h_\pm^{(j)}(x_j,t_o)=e^{it_oH}h_\pm^{(j)}(x_j)e^{-it_oH}$. By chirality, $h_\pm^{(j)}(x_j,t_o)=h_\pm^{(j)}(x_j\mp t_o)$ with no discontinuity at the origin by the use of the boundary conditions associated to the $H$-dynamics\footnote{This applies to the hamiltonian densities for any topological defects but not to other chiral operators if the defect is non-trivial although topological.}. For any given $x_j,\,y_j$ there are $R\gg t_o$ large enough such that $x_j-t_o \in [-R/2,0]$ and $y_j+t_o\in[0,R/2]$, so that the left/right movers have been moved into the two sub-systems. There, the expectations (\ref{statvev}) factorize and are equal to 
\[
 \bra \prod_j h^{(j)}_+(x_j-t_o)\ket_0^l\bra \prod_k h^{(k)}_-(y_k+t_o)\ket_0^r.
\]
Correlation functions of pure right-mover or of pure left-mover hamiltonian densities are translation invariants, and we can drop the $t_o$ dependence in the previous equation. Hence, the limit in eq.(\ref{statvev}) exists and the steady state factorizes on left/right movers as heuristically argued above. This factorization is found in the XY chain \cite{Pillet}, but our proof is valid for arbitrary gapless (critical) systems. It is simple to see that this result applies also to multi-time correlation functions. Note that there does not seem to be a geometrically simple Euclidean field theory description of the resulting stationary measure, contrary to equilibrium finite-temperature. Before the connection we have two semi-infinite cylinders of circumferences $\beta_l$ and $\beta_r$, but after the connection and an infinite real time evolution, we find a separation between right- and left-movers, hence no immediate Euclidean space geometry.

{\it Energy current.} For convenience, the quantity we choose to measure is the energy difference in the left and right sub-systems: $Q(t):=\frac{1}{2}(H^l(t)- H^r(t))$ with $H^l(t) = \int_{-R/2}^0 dx\, h(x,t)$ and $H^r(t) = \int^{R/2}_0 dx\, h(x,t)$ evolved in time with the $H$-dynamics with reflecting boundary conditions at $\pm R/2$. Since by chirality $h_\pm(x,t) = h_\pm(x\mp t)$ with the interpretation that through $\pm R/2$ they are interchanged thanks to the reflection, we find $H^l(t)=H^l +\int_0^t dx\, (h_-(x)-h_+(-x))$ and similarly for $H^r(t)$. Thus,
\begin{eqnarray}\label{defQt}
Q(t)=Q  +\int_0^t dx\, (h_-(x)-h_+(-x)),
\end{eqnarray}
and the mean energy current is $J=\bra h_+(-t)-h_-(t)\ket_{\rm stat}$. Eq.(\ref{defQt}) has a simple interpretation: the energy transferred during time $t$ and its statistic only involve the hamiltonian densities at distances at most $t$ from the contact point since the latter propagate uniformly at constant velocity. Factorization of the stationary measure gives $J= j(\beta_l) - j(\beta_r)$. The function $j(\beta)$ can be computed by modular transformation, following arguments used in studying finite size effects \cite{finitesize}. The results is $j(\beta) ={\pi c}/{12 \beta^2}$, so that $J = \frc{\pi c}{12} (\beta_l^{-2} - \beta_r^{-2})$ as announced in eq.~(\ref{heatJ}).

{\it Energy transfer statistics.}
Let us now turn to the energy FCS (\ref{energyFCS}). One has to be careful on how to define the energy transfer during time $t$. We assume a two-step measurement process: Once the stationary regime has been reached, first the energy difference $Q$ is measured at time $0$. The output is $q_0$ with probability ${\rm Tr}(P_{q_0}\rho_{\rm stat})$, where $P_{q_0}$ is the projector on the corresponding eigenspace. Then, at later time $t$, $Q$ is again measured. The output is $q$ with probability $P_t(q,q_0)={\rm Tr}(P_{q}e^{-itH} P_{q_0}\rho_{\rm stat}P_{q_0} e^{itH}P_q)$. The heat transfer generating function is defined as 
	\[ \bra e^{i\lambda \Delta_tQ}\ket:= \sum_{q,q_0}e^{i\lambda(q-q_0)}\, P_t(q,q_0) .\]

Since $Q$ has a discrete spectrum at finite $R$, this sum can be dealt with \cite{Espo} using the formula $\int d\mu \,e^{i\mu(Q-q)}\propto P_q$, with an appropriate integration range; we also use the formula $\sum_q f(q) P_q = f(Q)$. This yields an integral representation $\bra e^{i\lambda \Delta_tQ}\ket\propto \int d\mu\, {\cal Z}_t(\lambda,\mu)$ with 
\begin{equation}\label{Zt}
{\cal Z}_t(\lambda,\mu):=\bra e^{-i\lt(\frc\lambda2-\mu\rt) Q} e^{i\lambda Q(t)}e^{-i\lt(\frc\lambda2+\mu\rt)Q}\ket_{\rm stat}\end{equation}
where $Q(t)$ is defined in eq.~(\ref{defQt}). Although the operator averaged in eq.~(\ref{Zt}) appears non-local, the Baker-Campbell-Hausdorff formula guarantees that only $Q(t)-Q$ and its evolution under $e^{i\kappa Q}$ for $\kappa$ finite (in a range determined by $\lambda$ and $\mu$) are actually involved. These are all finitely supported, whence the stationary limit (\ref{statvev}) exists and is described by the invariant measure. As is usual in this context \cite{Espo,BD11}, one expects the large-time limit of $Z_t(\lambda,\mu)$ to be $\mu$-independent, so that we may specialize to $\mu = \lambda/2$ for simplicity. The large deviation function is then
\begin{eqnarray} \label{Fdef}
F(\lambda) = \lim_{t\to\infty} t^{-1}\log {\cal Z}_t(\lambda,\lambda/2).
\end{eqnarray}

Using the construction of the invariant measure, we find, as expected, the factorized expression 
\begin{eqnarray}\label{factoF}
F(\lambda)={f(\lambda,\beta_r) + f(-\lambda,\beta_l)}
\end{eqnarray}
with 
$ f(\lambda,\beta,t):= \lim_{t\to\infty} t^{-1}\, \big(\lim_{R\to\infty}\log \bra G_\lambda(t) \ket_\beta\big),$
where  the expectation $\bra\cdots\ket_\beta$ is taken in the CFT on the interval $[0,R/2]$ at temperature $\beta^{-1}$. 
Here, $G_\lambda(t)$ is the chiral factor of $e^{i\lambda Q(t)} e^{-i\lambda Q}$. The CFT computation of these expectations and their large time limits will be detailed in \cite{BD12}, it leads to eq.~(\ref{energyFCS}). 
Instead, we here present a simpler derivation of eq.~(\ref{energyFCS}),  following only from the fluctuation relation, the above factorization (\ref{factoF}), scale invariance, and the leading small-$\lambda$ asymptotic behavior. The fluctuation relation is of course a consequence of the CFT computation that will be presented in \cite{BD12}, but here we must assume it. Let $z=i\lambda$. By scale invariance, $f(z,\beta)=z^{-1}g(z/\beta)$. On $g$, the fluctuation relation (\ref{fluctu}) translates into 
\[ 
g\big(\frac{u-v}{u+1/2}\big)-g\big(\frac{v-u}{v-1/2}\big)=(u-v)\big[g\big(\frac{1}{u+1/2}\big)-g\big(\frac{-1}{v-1/2}\big)\big],
\]
where we set $u+1/2=\beta_r/z$ and $v-1/2=\beta_l/z$. Expanding this equation to leading order in $(u-v)$ leads to
\[ g'(w)=\frac{c\pi}{12\hbar}\big[\frac{1}{(1-w)^2}-1\big],\]
where we set $u-1/2=-1/w$ and use $g(w)=\frac{c\pi}{12\hbar}w^2 +O(w^3)$, which comes from the mean current formula (\ref{heatJ}). Integrating this equation gives eq.~(\ref{energyFCS}).

{\it A classical Poissonian interpretation.}
The heat FCS (\ref{energyFCS}) possesses a natural interpretation in terms of classical Poisson process\footnote{This is similar to the Levy-Kintchin decomposition although $F$ is a large deviation function and not the characteristic function of an infinitely divisible process.}. Observe first that $F(\lambda)=F^r(\lambda)-F^l(-\lambda)$ can be decomposed as
\begin{eqnarray} \label{Fpoisson}
 F^{l,r}(\lambda)= \int d\nu^{l,r}(\varepsilon)\, (e^{i\lambda \varepsilon}-1) 
 \end{eqnarray}
with measure $d\nu^{l,r}(\varepsilon)=\frac{c\pi}{12\hbar}\,e^{-\beta_{l,r}\varepsilon}\,d\varepsilon$ for $\varepsilon>0$ and $0$ otherwise, so that $F(\lambda)$ coincides with the generating function of the difference of time-homogeneous Poisson processes with intensity $d\nu^{l,r}(\varepsilon)$, that we denote by ${\cal E}_t$: $\mathbb{E}[ e^{i\lambda {\cal E}_t} ] = \exp[t F(\lambda)]$. Recall that a Poisson process is a piecewise constant but discontinuous stochastic process whose jumps are Poisson variables. Alternatively, the energy transfer $d{\cal E}_t$ during time $dt$ may be represented as the sum of its jumps, that is $d{\cal E}_t=\int \varepsilon[dN^r_t(\varepsilon)-dN^l_t(\varepsilon)]$ where the numbers $dN^{l,r}_t(\varepsilon)$ of jumps of size in $[\varepsilon,\varepsilon+d\varepsilon]$ during time $dt$ are independent Poisson variables with mean $d\nu^{l,r}(\varepsilon)dt$.
The representation (\ref{Fpoisson}) of the FCS has a simple interpretation. A jump of ${\cal E}_t$ of size $\varepsilon<0$ (resp. $\varepsilon>0$) corresponds to an energy quanta transfer from left to right (resp. from right to left). Transfers of particles occur (without scattering) randomly homogeneously and independently in time with a probability proportional to the Boltzmann weight $e^{-\beta_{l,r}\varepsilon} d\varepsilon dt$, i.e. particles transferring from left to right (resp. from right to left) have been prepared with temperatures $\beta_l^{-1}$ (resp. $\beta_r^{-1}$). 

This representation, which applies to CFT energy full counting statistics, leads to possible conjectural generalizations which are all based on assuming that the FCS is that of Poisson processes of particle transfers.
The first consists in including charge transfer counting statistics (without scattering) by assuming that the energy quanta also carry charges; we will develop this in a later work.
The second consists in considering cases in which particle energy densities may not be flat as for 1d massless particle. 
To take this into account we are tempted to conjecture that the intensities of the processes should be modified according to: 
\[\propto  \frac{d\varepsilon dt}{\hbar}\, \exp({-\beta_{l,r}(\varepsilon-T_{l,r}s(\varepsilon))}),\]
where $s(\varepsilon)$ is the entropy, so that $e^{s(\varepsilon)/k_B}$ is the degeneracy of energy $\varepsilon$. This should apply to 1d gapped systems with particles with dispersion relation $\varepsilon(p)^2=p^2v_f^2+m^2v_f^4$ with $\Delta=mv^2_f$ the energy gap and $m^{-1}$ the band curvature at the gap. In 1d, the degeneracy is flat in momentum space, and the intensities would be $\propto \frac{v_f}{2\pi\hbar}\,{dpdt}\,e^{-\beta_{l,r}\varepsilon(p)}$. This may easily be generalized to higher dimensions.

{\it Comments.}
Eq.~(\ref{energyFCS}) provides elements of information on non-equilibrium dynamics and energy transfers in CFT. The result is very universal: it only depends on the universality class of the critical point, and further, only on one parameter characterizing this universality class, the central charge. Its derivation in \cite{BD12} will further give a check of the fluctuation relations in non-trivial quantum interacting systems. It is also worth noticing that the universal mean energy current (\ref{heatJ}) and its fluctuations are independent of the excitation velocity $v_f$. So putting CFT out of equilibrium provides a way to determine, numerically or experimentally, its central charge free of non-universal unknown parameters. Generalizing the above results to cases with non-trivial defects  partially reflecting the energy (i.e.~with an energy transmission coefficient $|t|\neq1$) and/or with sub-systems described by two different CFT would be interesting \cite{BD12}, as would be generalizations to integrable models\footnote{Using either thermodynamic Bethe ansatz or known methods for evaluating one-point averages.}. The representation (\ref{Fpoisson}) applies nicely to FCS of commuting charges but its generalization to FCS of different non-commuting charges remains a mystery.

{\it Acknowledgements}: DB thank M. Bauer and J. Cardy for discussions, and BD thanks F. Essler, G. Watts and R. Weston for sharing ideas. This work was in part supported by ANR contract ANR-2010-BLANC-0414.

\end{document}